\documentclass[twocolumn,12pt,cites,graphicx]{article}
\usepackage{epsfig}
\usepackage{psfrag}
\usepackage{subfigure}

\title{Screening in two-dimensional foams}
\author{
S.J. Cox$^{\rm a,b}$  \and F. Graner$^{\rm b}$ \and M.F. Vaz$^{\rm c}$ \\[2ex]
$^{\rm a}$ Institute of Mathematics and Physics, Aberystwyth University, \\
 Ceredigion SY23 3BZ, United Kingdom, Email: foams@aber.ac.uk\\
$^{\rm b}$ Laboratoire de Spectrom\'{e}trie Physique, UMR5588, \\
CNRS-Universit\'{e} Grenoble I, B.P. 87, F-38402 Martin d'H\`{e}res Cedex, France\\
$^{\rm c}$ Instituto de Ci\^encia e Engenharia de Materiais e Superficies and \\
 Departamento de Engenharia de Materiais, \\
Instituto Superior T\'ecnico, Avenida Rovisco Pais, 1049-001 Lisboa, Portugal
 }

\date{\today}

\begin{document}

\maketitle

\renewcommand{\thefootnote}{\fnsymbol{footnote}}

\noindent
Using the Surface Evolver software,
we perform numerical simulations  of point-like deformations in a two-dimensional foam.
We study perturbations which are infinitesimal or finite, isotropic or anisotropic, and we either conserve or do not conserve the number of bubbles. We measure the  displacement fields around the perturbation. Changes in pressure decrease exponentially with the distance to perturbation, indicating a screening over a few bubble diameters.


\section{Introduction}
\label{sec:intro}

A foam is a discrete material, made of gas bubbles separated by a continuous liquid phase.  Foams act as elastic solids for small deformations, but when 
large strains are applied they behave as viscous liquids; applied stresses are relaxed by discrete rearrangement events that occur in the foam. Other changes are due to ageing, where some bubbles gain gas at the expense of others.
In most cases of foam evolution, continuous changes, such as infinitesimal changes of bubble shapes and sizes, alternate with discontinous processes
\cite{glazierw92}.

For instance, the length of an edge might decrease (or, inversely, it might stretch). If it  vanishes, a neighbour-swapping event occurs, and a new edge is created: this is the topological  T1 process  \cite{weairer84} (its inverse is a T1 too). Alternatively, an edge breakage leads to bubble coalescence, also called fusion (its inverse is a division). 
All these perturbations are anisotropic. On the other hand, when a bubble's area decreases, it is an isotropic pertubation (its inverse is a bubble inflation). If it vanishes, a reduction in the number of bubbles occurs: this is the topological process T2  \cite{weairer84}, reminiscent of biological cell apoptosis (its inverse is a nucleation).
 
Each pertubation affects the neighbouring bubbles, over a certain range. This range has been the subject of various studies, with diverse motivations, all in two dimensions. 
The perturbation induced by a T1 has been measured in experiments 
  \cite{eliasbms99} and in simulations \cite{kablad03}.
  The effect of changing the volume of a single bubble is also studied, both in 
experiment \cite{kader97,guene} and in simulations \cite{levitan,mancini}.
A laser has been used to break the wall between bubbles   \cite{lordereau}. 
This laser ablation is also used in biological tissues
  \cite{fahadifar07}, and cell division and apoptosis  are currently studied too \cite{courtypreparation}, motivated by a desire to understand the mechanics of biological tissue.
  
An infinitesimal change is expected to have an infinitesimal effect on neighbouring bubbles.  
Conversely, finite discontinuities can have a finite effect on other bubbles.
If these events are scarce enough, and spatially screened by the foam disorder, their effects on the foam are independent, and the foam can at large scale be treated as a continuous material \cite{dollet07}. 
 The foam properties of  insulation against sound and explosion probably derive from this screening.
A question under debate is to determine under which conditions these processes correlate, triggering for instance a cascade of T1s \cite{glazierw92,eliasbms99,guene,mancini}.
This can 
lead to large-scale fluctuations such as avalanches \cite{rosaf98,tewari,cohen2001,lauridsentd02}.
The role of disorder on perturbation screening is also of potential interest in various other discrete systems
 \cite{picard04,maloneyl06,Tanguy2006}. 

To study the screening in foams, we focus on a 2D simulated foam with an isolated perturbation. 
   Physically, this is relevant for a system which is 
initially formed by well separated defects, that may eventually interact. We try to distinguish a screening area around the perturbation, and the rest of the foam where no disturbance is felt.
   
We simulate small continuous geometrical changes, then finite discontinuous topological processes.
We either conserve or change the number of bubbles. We try different foam disorders and boundary conditions.
We perform isotropic (scalar) pertubations, and anisotropic (tensorial) ones. 
Two types of measurements are extracted from the simulations:  the displacement of each bubble (a vector), and the  change in pressure  \cite{fortesmv07}  in each bubble, with respect to the inital (unperturbed) foam.

\section{Method}

We create a disordered foam structure using a Voronoi construction based upon a random Poisson process for generating the cell centres \cite{brakke86}. The cell areas $A$ are chosen based upon a randomly generated Weibull distribution \cite{wyndc08}:
\begin{equation}
f(A; \beta, \lambda) =  \frac{\beta}{\lambda} \left( \frac{A}{\lambda}\right)^{\beta-1}  e^{-({A}/{\lambda})^\beta},
\end{equation}
where the parameter $\beta > 1$  determines the area dispersity and the parameter $\lambda$ is chosen as $\lambda = 1.115 \langle A \rangle$, so that the peak of the distribution is close to $A = \langle A \rangle$. For the monodisperse case, each cell is set to have the same target area, rather than taking the limit $\beta \rightarrow \infty$. We also constructed ordered monodisperse foam consisting of 8 and 20 concentric shells of hexagons \cite{coxg03}.

The disorder of each foam is defined by the (dimensionless) variance of the area distribution:
\begin{equation}
var(A) = \left\langle 
\frac{(A - \langle{A}\rangle)^2}{\langle A \rangle^2}
\right\rangle
\label{eq:mu2A}
\end{equation}
where $\langle \rangle$ denotes an average over the whole foam.

\begin{figure}
\centerline{
\includegraphics[width=10cm]{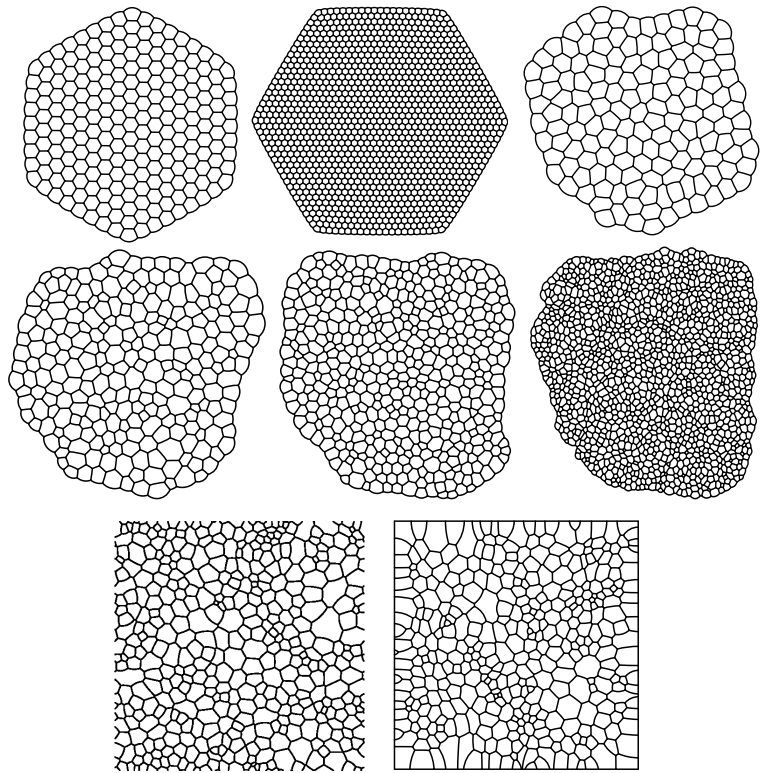}
}
\caption{Representative images of the foams used. Top row: Monodisperse foams, both ordered (217 and 1261 bubbles) and disordered (150 bubbles). Middle row: Polydisperse foams, in each case with almost the same value of polydispersity, $var(A) \approx 0.12$, and 250, 450 and 1400 bubbles respectively. Bottom row: Polydisperse foams without boundaries (periodic, 400 bubbles) and with boundaries (383 bubbles) respectively.}
\label{fig:foams}
\end{figure}

Each foam is first equilibrated in the Surface Evolver software \cite{brakke92}, using a mode in which each film is represented exactly as a circular arc and a value of surface tension $\gamma$ equal to one. T1 topological changes are automatically triggered when films shrink to a very small length during the equilibration process. Some of the initial foam structures used are shown in figure \ref{fig:foams}.

We then trigger changes to the structure as follows, illustrated in figure \ref{fig:subpics}:
\begin{itemize}
\item An {\em edge stretch} is performed on the longest edge of the bubble nearest the centre of the cluster, by extending or reducing its length by a fraction $\delta L$ and fixing its endpoints.
\item An {\em inflation} of the bubble nearest the centre of the cluster by a fraction $\delta A$.
\item A {\em T1} is performed on the shortest edge of the bubble nearest the centre of the cluster.
\item A {\em T2} event is triggered by removing the bubble nearest the centre of the cluster.
\item A {\em coalescence} event is triggered by removing the longest edge of the bubble nearest the centre of the cluster.
\item A {\em division} event is triggered by inserting a edge across the centre of the bubble nearest the centre of the cluster, giving two bubbles each with half the area.
\end{itemize}

In each case we re-equilibrate the structure; we allow T1s to occur, but in fact they occur infrequently, except after the T2. We record the position of the bubble centres (average of the vertex positions) and the bubble pressures before and after the distortion, denoted $x_i$, $y_i$, $p_i$ with $i = 0,1$.

\section{Bubble displacement}


For free clusters, 
the displacements of the bubble centres after each event are shown in figure \ref{fig:displacement}. Since we apply no constraint during the equilibration to prevent rigid body displacements and rotations, we first calculate the average over all bubble centres of the $x$ and $y$ displacement and the average rotation about the centre of the cluster. These averages are then subtracted from the displacement of each bubble. The results suggest that T1s, T2s, coalescence, division and edge-stretching events all induce approximately quadrupolar displacement fields, while the inflation event is almost purely radial.
It is surprising to see that the effect of the T2 is not isotropic, although not surprising to see that it has the largest effect (greater displacements).

This information is summarised in figure \ref{fig:deltar}, which shows
the variation in position of each bubble centre as a function of radial and angular position. 
The radial displacement decreases with radial distance, and, apart from the inflation event, varies sinusoidally with angle like a quadrupole, $\Delta r \sim \sin(2\theta)$.
The angular displacement field is slightly asymmetric but also approximately sinusoidal and dipolar, $r_0 \Delta \theta \sim \sin(\theta)$.
The exception is the inflation event, for which the motion is radially outwards.

This dipolar angular displacement is not screened and is sensitive to boundary conditions. 
In foams with periodic boundary conditions the angular displacement deviates from a sinusoid (figure \ref{fig:boundaries}). In foams with fixed boundaries  (figure \ref{fig:foams}, bottom), which force the radial displacement to vanish (equivalently, each pertubation interacts with its reflection at boundaries, or virtual  image), the angular displacement becomes quadrupolar.
 
\section{Pressure}

The perturbation induces a change in each bubble pressure, $\Delta p = p_1-p_0$, with respect to the initial foam.
 We normalize it by $\gamma/\sqrt{\langle A \rangle}$ to facilitate comparison between different foams. 
It  decreases with radial distance $r_0 = \sqrt{x_0^2+y_0^2}$, as shown in figure \ref{fig:randmonopr}.
 
A small change in the area of a bubble at the centre of the cluster of 150 bubbles, in this case by a factor of 20\%, has only a small effect on the pressure differences, which decrease roughly exponentially with radial distance. A similar result was found in a cluster of 1400 bubbles.

All other distortions induce a stronger variation in pressure difference. 
The largest pressure differences are associated with the T2 and coalescence events.
In the case of a change in edge length, there is little difference between extension and compression;
similarly, inflating and shrinking a bubble have the same effect.

The pressure change fluctuates from bubble to bubble, and is either positive or negative. 
Its standard deviation visibly decreases (figure \ref{fig:randmonopr}). 
In an ordered cluster, it is possible to define radially concentric shells of bubbles and examine the pressure change in each cell, but to quantify the pressure differences in disordered foams the shells vary greatly in composition \cite{astebr96} and it is necessary to bin the data in some manner.
We choose a bin width of one bubble diameter $D=2\sqrt{\langle A \rangle /\pi}$, and increase the position of the centre of the bin in intervals of $D/10$. The average pressure difference $\Delta p$ is calculated 
 in each bin: it is always close to zero beyond the central bubble. 
The normalised standard deviation in each bin, $std(\Delta p)$, directly compares the effect of the different pertubations we apply (figure \ref{fig:outer}).

Remarkably, all pertubations decrease with a similar  exponential decay, sometimes over more than two decades. Its characteristic length, or screening length $\ell_s$, is the inverse of the slope:
\begin{equation}
\frac{1}{\ell_s} = \frac{\partial \log\left(std(\Delta p)\right)}{ \partial {r}} . 
\label{eq:slope}
\end{equation}
 Although we did not find quantitative measurements in the literature, the available qualitative data  
\cite{eliasbms99, fahadifar07,courtypreparation} 
seem to observe a screening over one or a few bubble diameters. 
This is what we observe for the different distortions
(figure \ref{fig:alldata2}): despite large changes in several parameters, there is little variation in $\ell_s$, with no clear dependence upon cluster size (data not shown) or disorder. It is in general larger for confined foams and those with periodic boundary conditions. 

The choice of $var(A)$ as the disorder parameter requires some discussion: since the screening is a local effect, we should ideally use a local measure of disorder. It is, however, not clear how to define such a measure.

For all distortions except the bubble inflation, the  exponential decrease of $std(\Delta p)$ crosses over to a plateau  near the outer part of the foam.
The cross-over length $r_c$ marks the limit where the effect of the pertubation can be detected. 
We measure it as the intersection of linear fits to the decrease and the plateau (figure \ref{fig:outer}b).  Its value is very robust (figure \ref{fig:alldata1}): it
is roughly 60\% of the cluster radius independent of pertubation type and boundary conditions; as well as of  polydispersity and cluster size.
In the present simulations, it corresponds to roughly  5 to 15 bubble diameters.
Thus, we can change the volumes of bubbles around the remaining 40\% outer part of the foam:  it does not change the screening length (figure \ref{fig:outer}b).

\subsection{Other scalar measures of screening}

The perturbation induces other changes in the structure of the foam, for example each edge may shrink or lengthen to accommodate changes at the centre of the cluster. We justify {\em a posteriori} our choice of pressure difference as a measure of screening by comparing it with (i) the change in each bubble's perimeter, $\Delta e$,  normalized by the square-root of its area,   and (ii) the change in the length of each edge $\Delta \ell$ \cite{eliasfggj99}, normalized by $\sqrt{\langle A \rangle}$. Figure \ref{fig:otherscalar} shows that neither of these scalar quantities indicates any screening in both ordered and disordered foams, while the pressure change does. In particular,  $\Delta \ell \sim r^{-2}$ over more than a decade. 

\section{Summary and perspectives}

In simulated 2D foams, we have performed continuous and discontinuous perturbations, either isotropic or anisotropic, which conserve or not the number of bubbles. We have varied  polydispersity  over a decade and cluster size  over almost a decade, with free, periodic or fixed boundary conditions.

The bubble displacements are dipolar in all cases, except for bubble inflation. They extend to the foam boundary and are thus sensitive to boundary conditions. 

The pressure change in each bubble fluctuates from one bubble to the other, even for bubbles the same distance from the perturbation. The standard deviation within each of these ``shells'' decreases exponentially with distance from the perturbation over up to two decades, with a characteristic (screening) length of the order of one average bubble diameter.

At a distance close to 60 \% of the foam radius,  the  standard deviation of pressure difference reaches a plateau. Beyond this distance the pertubation does not significantly affect the pressures.

Perspectives include the measurement of tensorial quantities such as the deformation; the comparison with displacement and deformation fields predicted by standard elasticity theory; direct comparison with experiments; vectorial anisotropic pertubations, such as obtained by moving a point-like defect within the foam (Stokes experiment  
\cite{dollet07}); and transition from low disorder to the perfectly ordered case (figure \ref{fig:monoord1261pr}).
 
\section*{Acknowledgements}

We thank  S. Courty,  E.H.M. Guene, E. Janiaud, J. K{\"a}fer, J. Lambert, M. Mancini and other participants in the Grenoble Foam Mechanics Workshop for stimulation and useful discussions. We thank K. Brakke for his development and maintenance of the Surface Evolver code. SJC thanks the British Council Alliance programme, CNRS and EPSRC (EP/D048397/1, EP/D071127/1) for financial support and UJF for hospitality during the period in which this work was conceived.


\clearpage \onecolumn

\begin{figure}
\centerline{
\includegraphics[width=10cm]{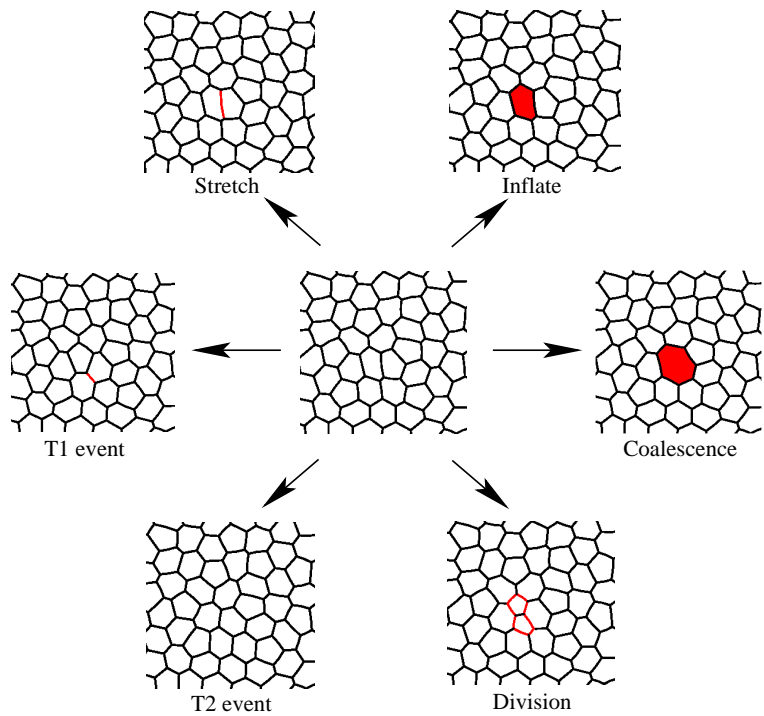}
}
\caption{The six types of distortion applied to the foams, illustrated here for the disordered monodisperse foam of 150 bubbles. Edge stretch (by 20\%), inflation (by 20\%), T1, T2, coalescence, division.}
\label{fig:subpics}
\end{figure}

\begin{figure}
\begin{center}
\mbox{
\subfigure[]{
\includegraphics[width=6cm,angle=270]{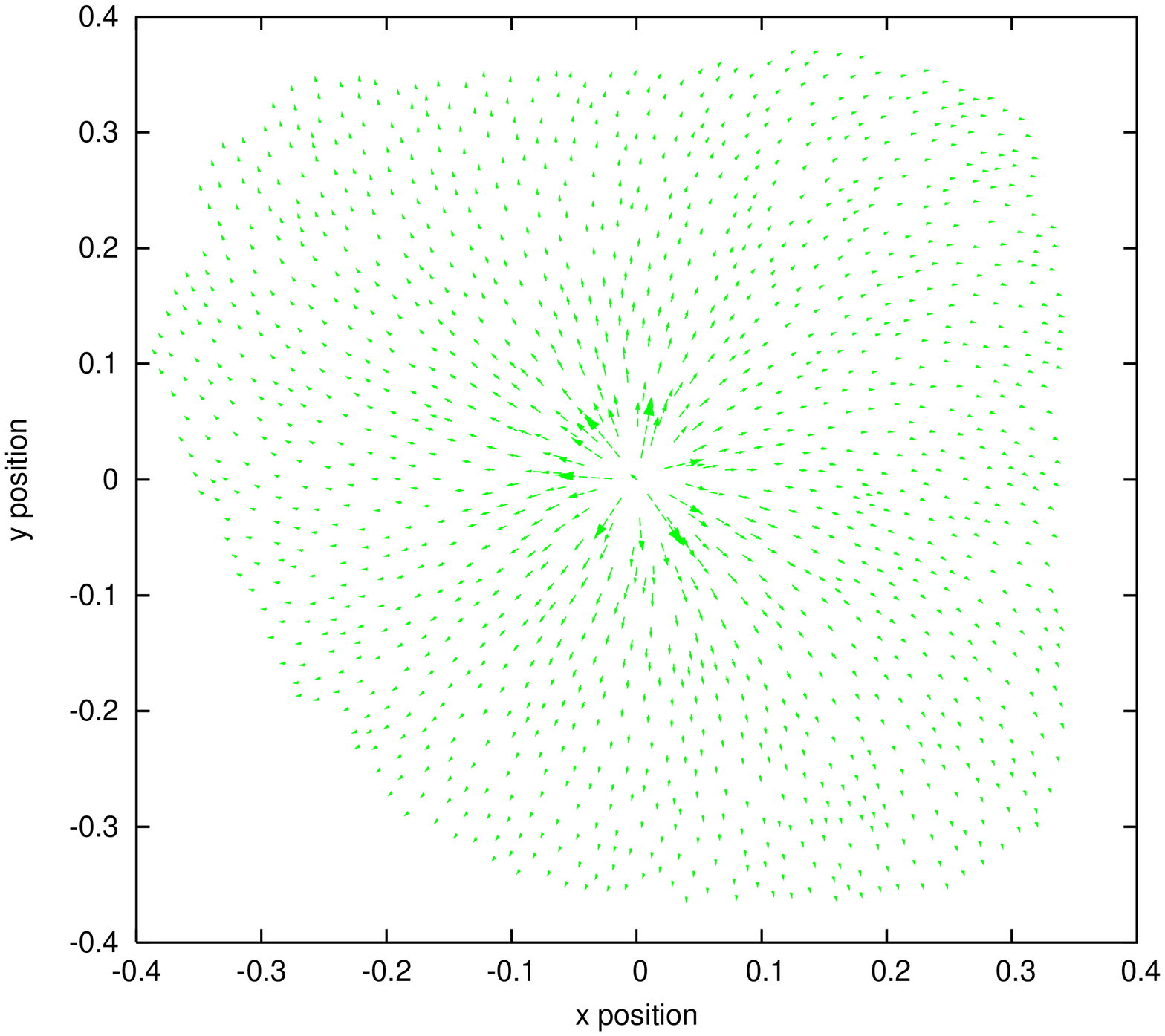}
}\quad
\subfigure[]{
\includegraphics[width=6cm,angle=270]{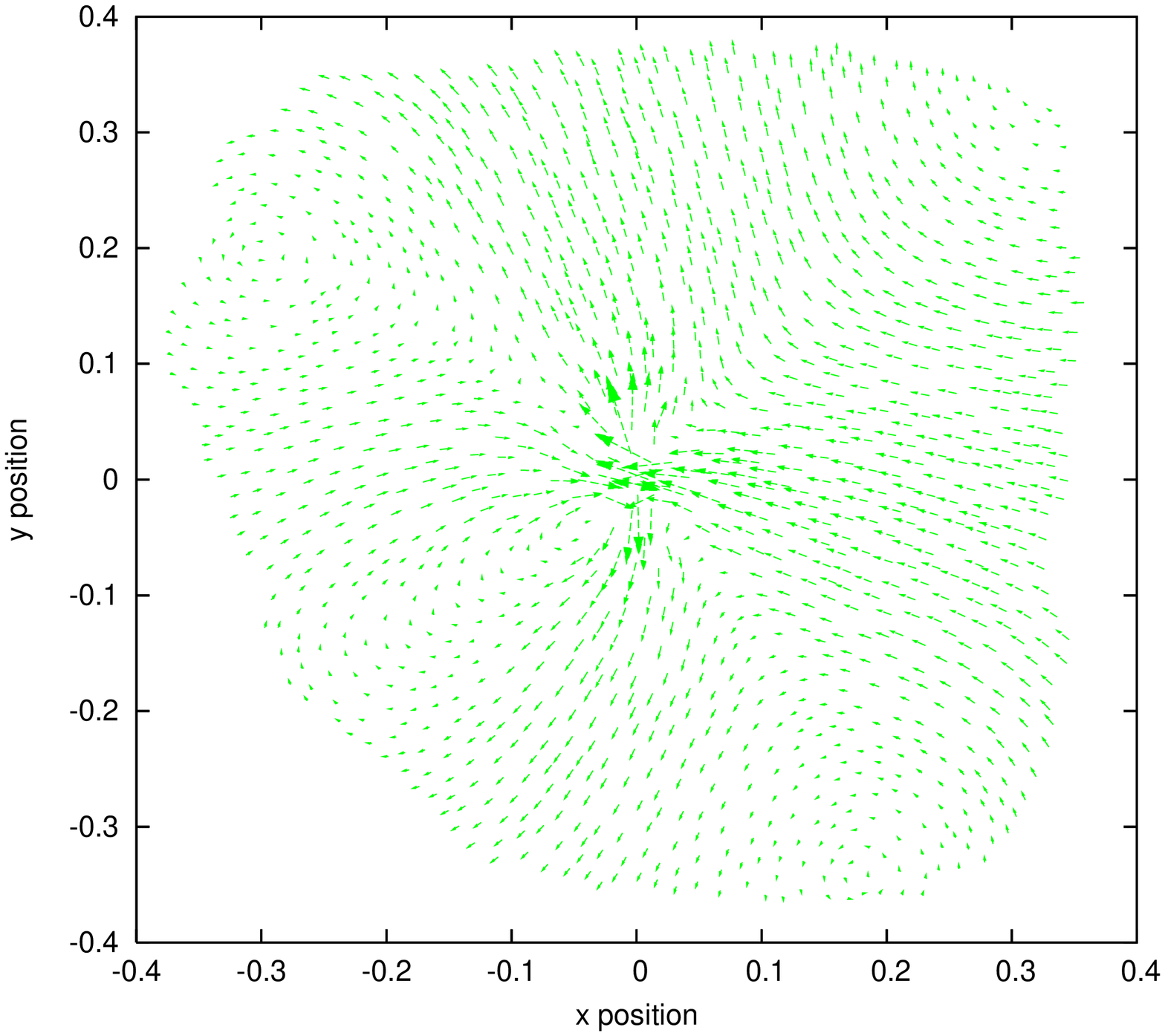}
}
}
\mbox{
\subfigure[]{
\includegraphics[width=6cm,angle=270]{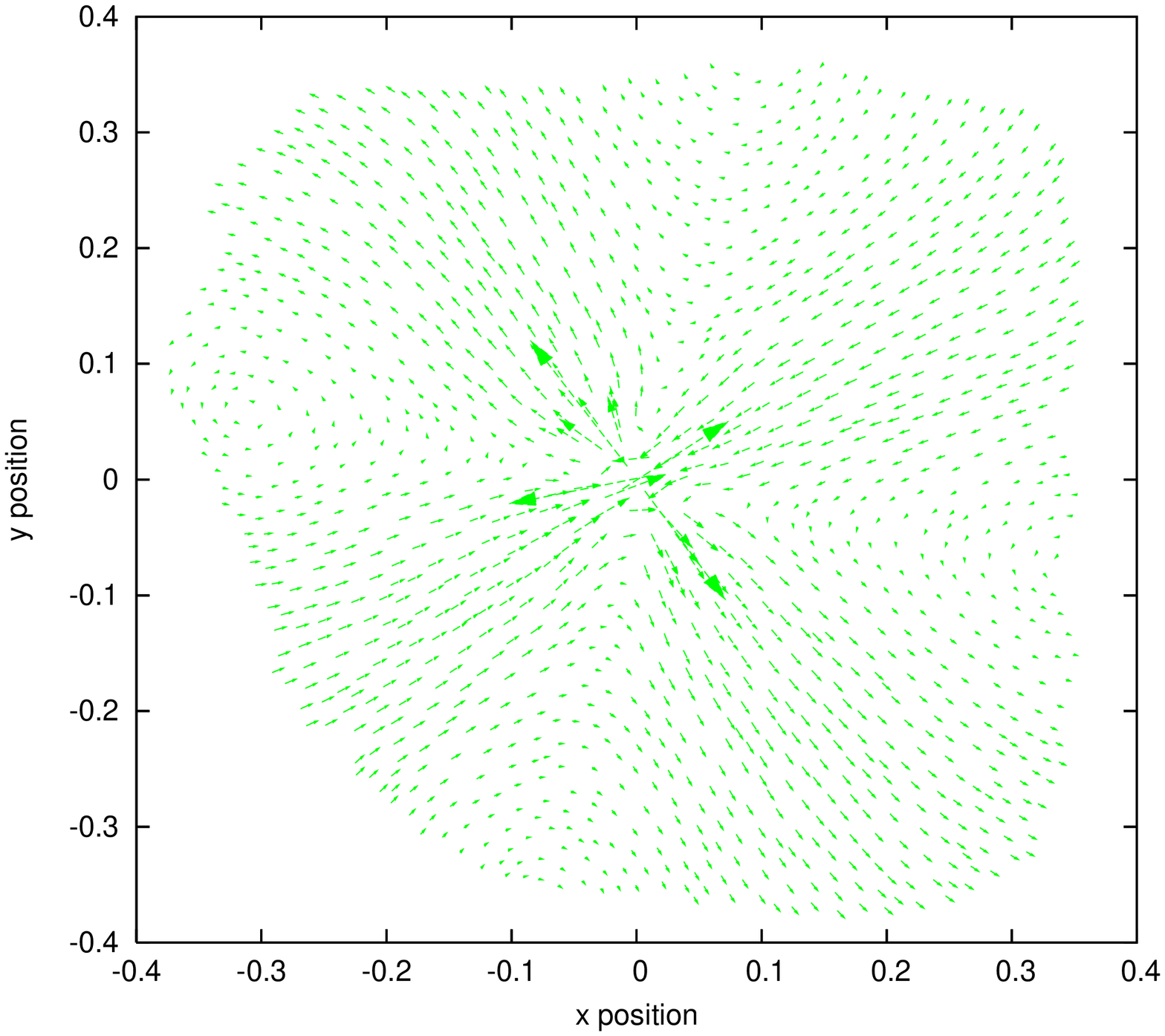}
}\quad
\subfigure[]{
\includegraphics[width=6cm,angle=270]{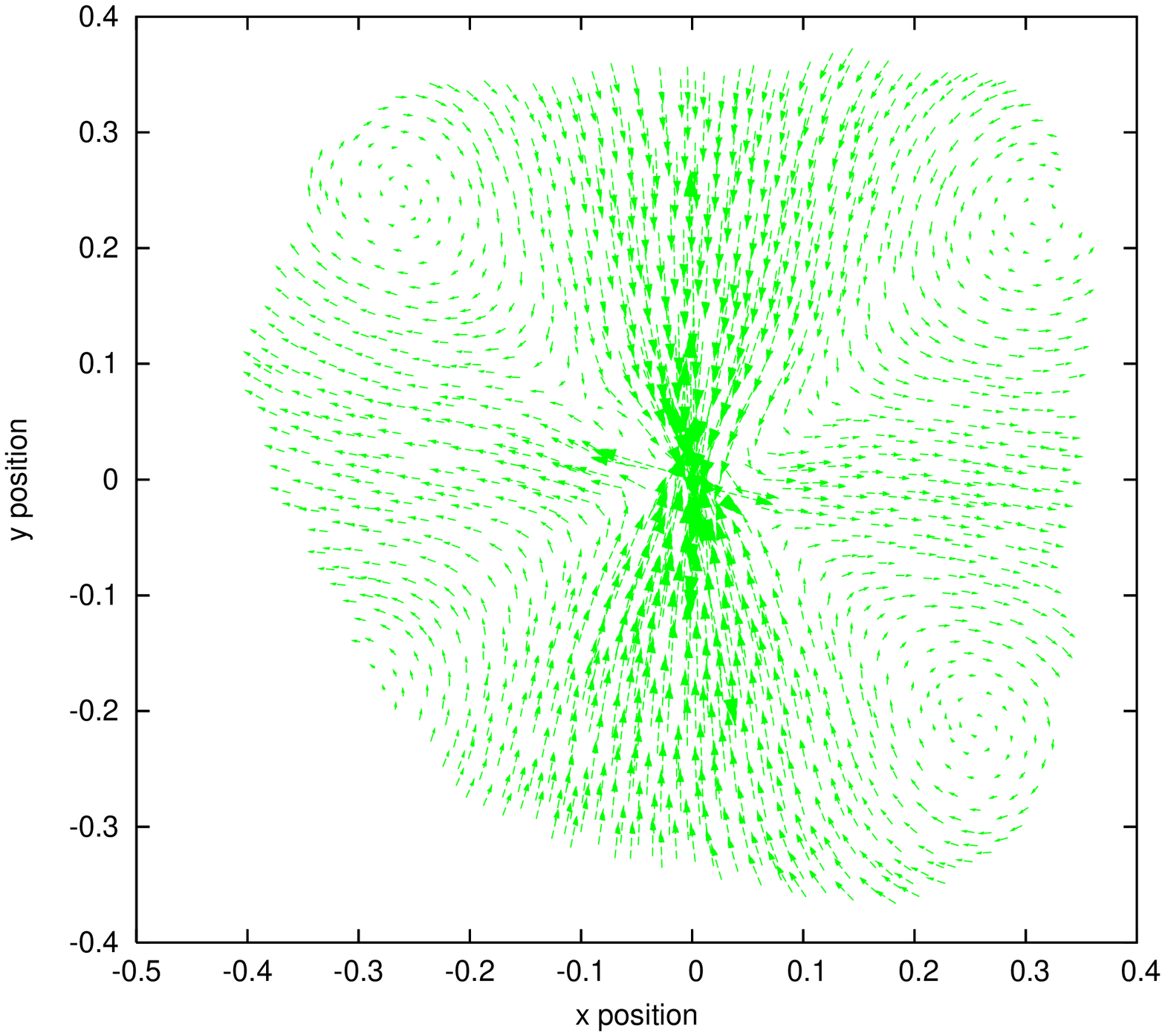}
}
}
\mbox{
\subfigure[]{
\includegraphics[width=6cm,angle=270]{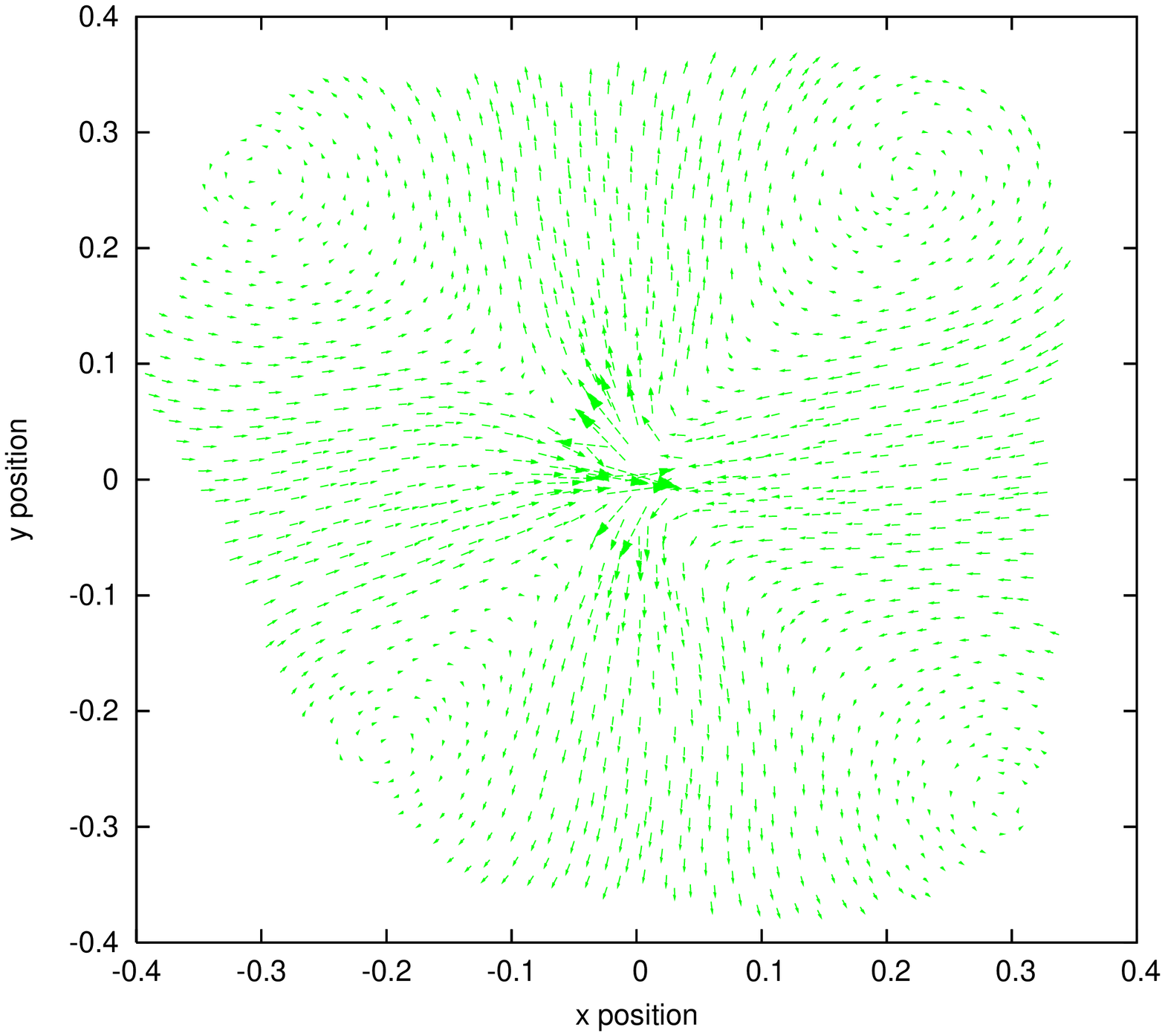}
}\quad
\subfigure[]{
\includegraphics[width=6cm,angle=270]{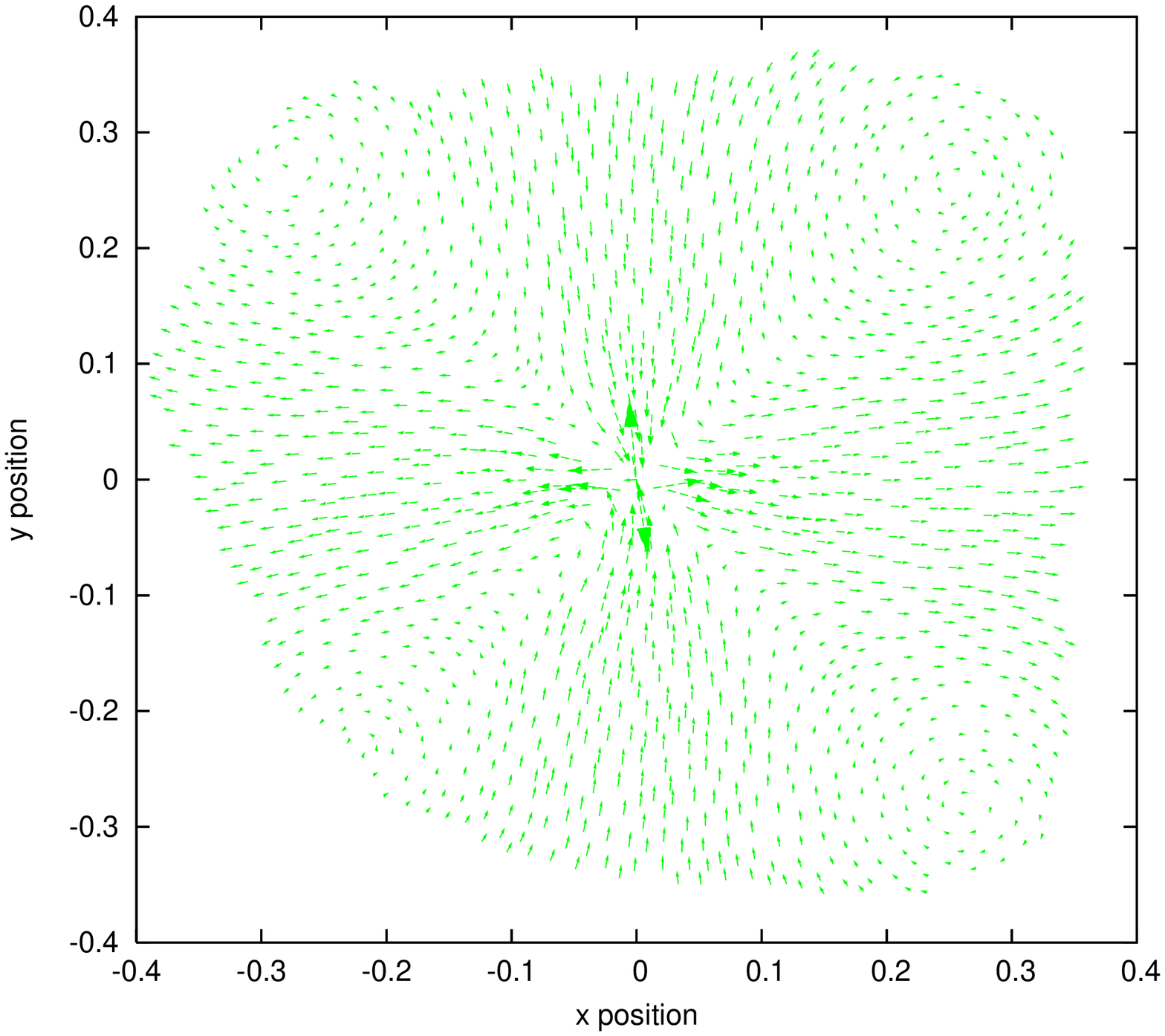}
}
}
\caption{The displacement field after (a) 10\% inflation, (b) 20\% edge stretch, (c) T1, (d) T2, (e) coalescence, and (f) division, in a cluster of 1400 bubbles with $var(A) \approx 0.12$. The average displacement and rotation has been subtracted in each case, and the vector length multiplied by a factor of 200 (in a), 100 (in b) or 30 (in c,d,e,f).}
\label{fig:displacement}
\end{center}
\end{figure}

\begin{figure}
\psfrag{Change in radial position}{Change in radial position, $\Delta r$}
\begin{center}
\subfigure[]{
\includegraphics[width=6cm,angle=270]{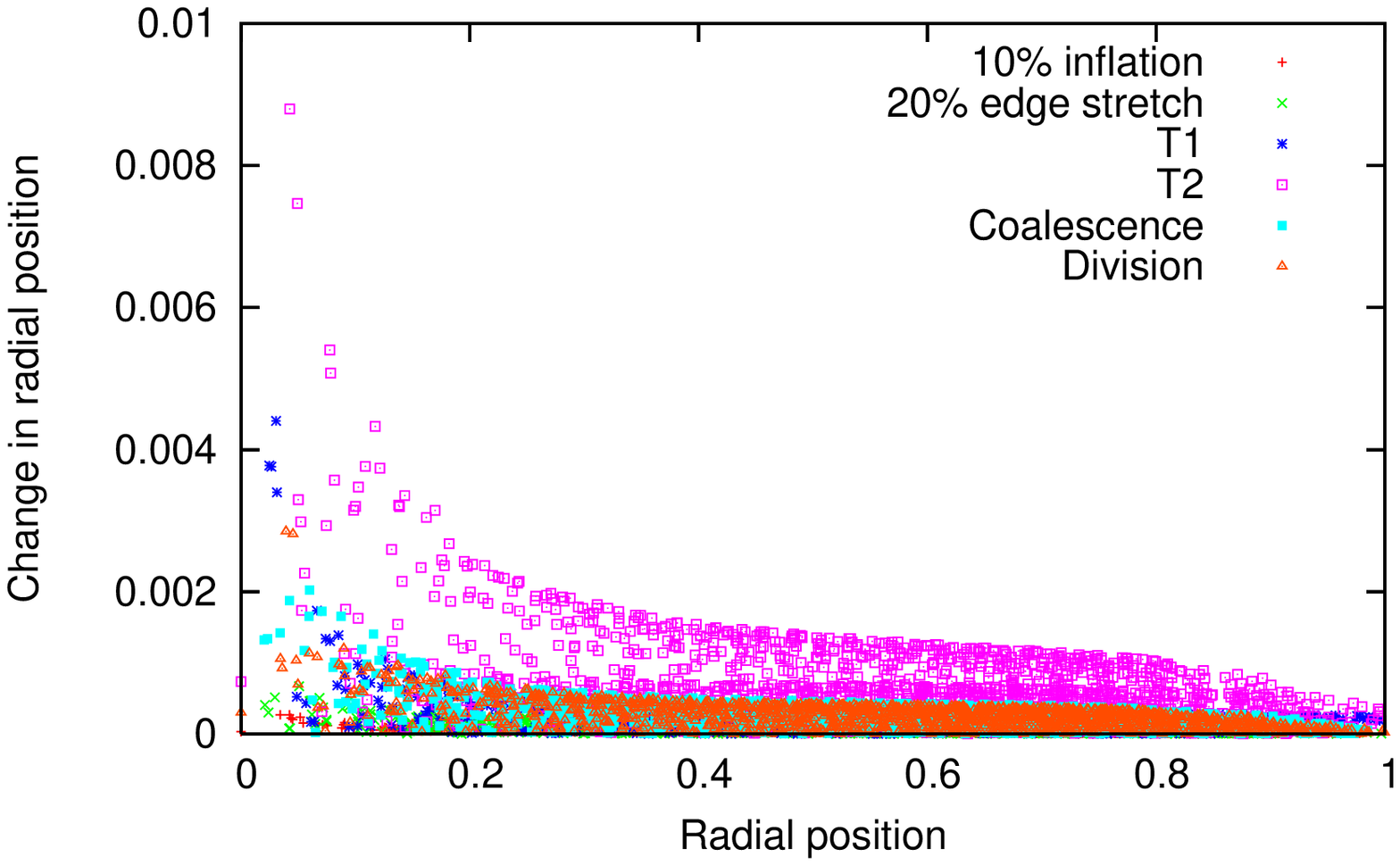}
}
\psfrag{Change in angular position, r0 dth}{Change in angular position, $r_0 \Delta \theta$}
\psfrag{Change in radial position}{Change in radial position, $\Delta r$}
\subfigure[]{
\includegraphics[width=6cm,angle=270]{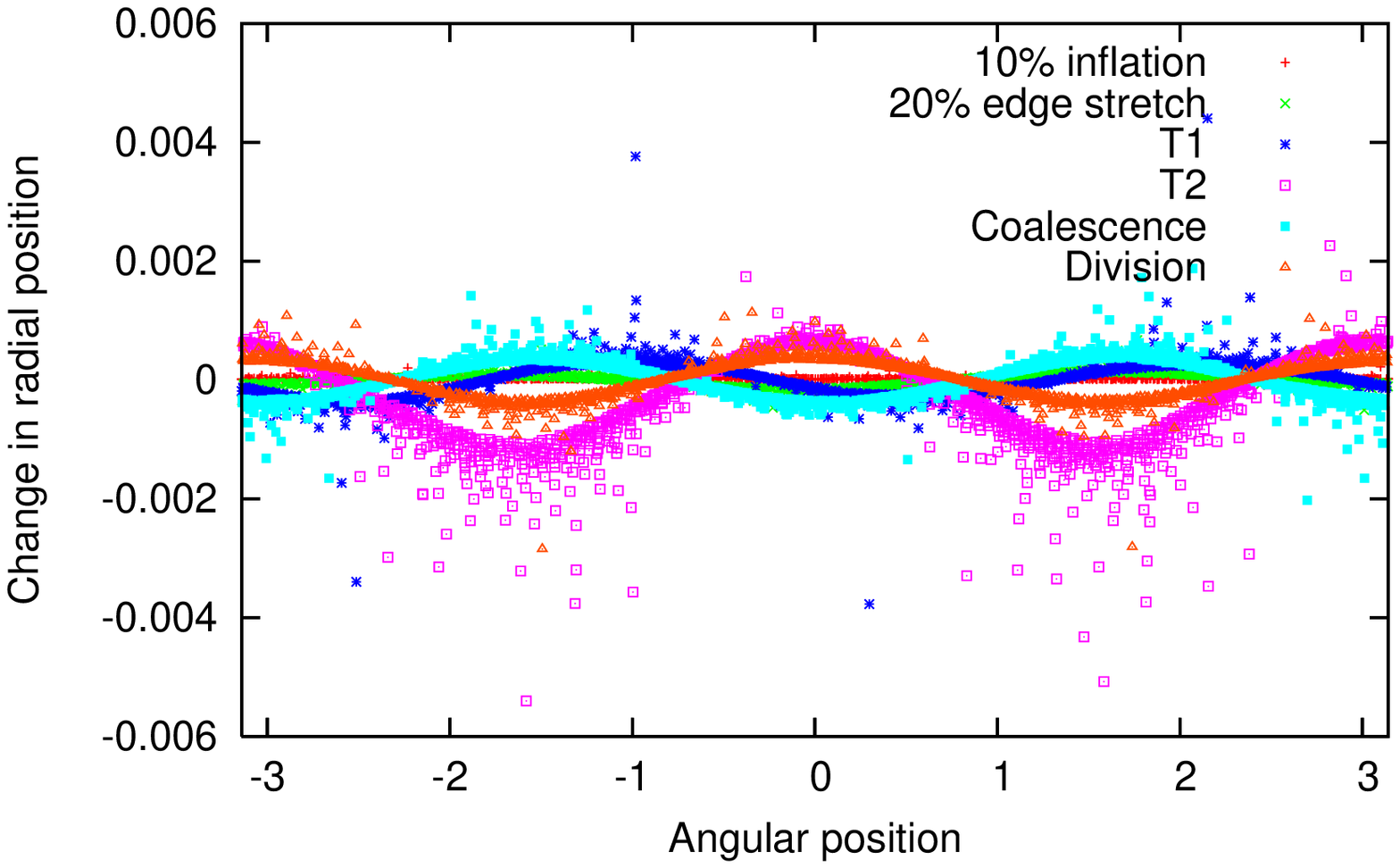}
}
\subfigure[]{
\includegraphics[width=6cm,angle=270]{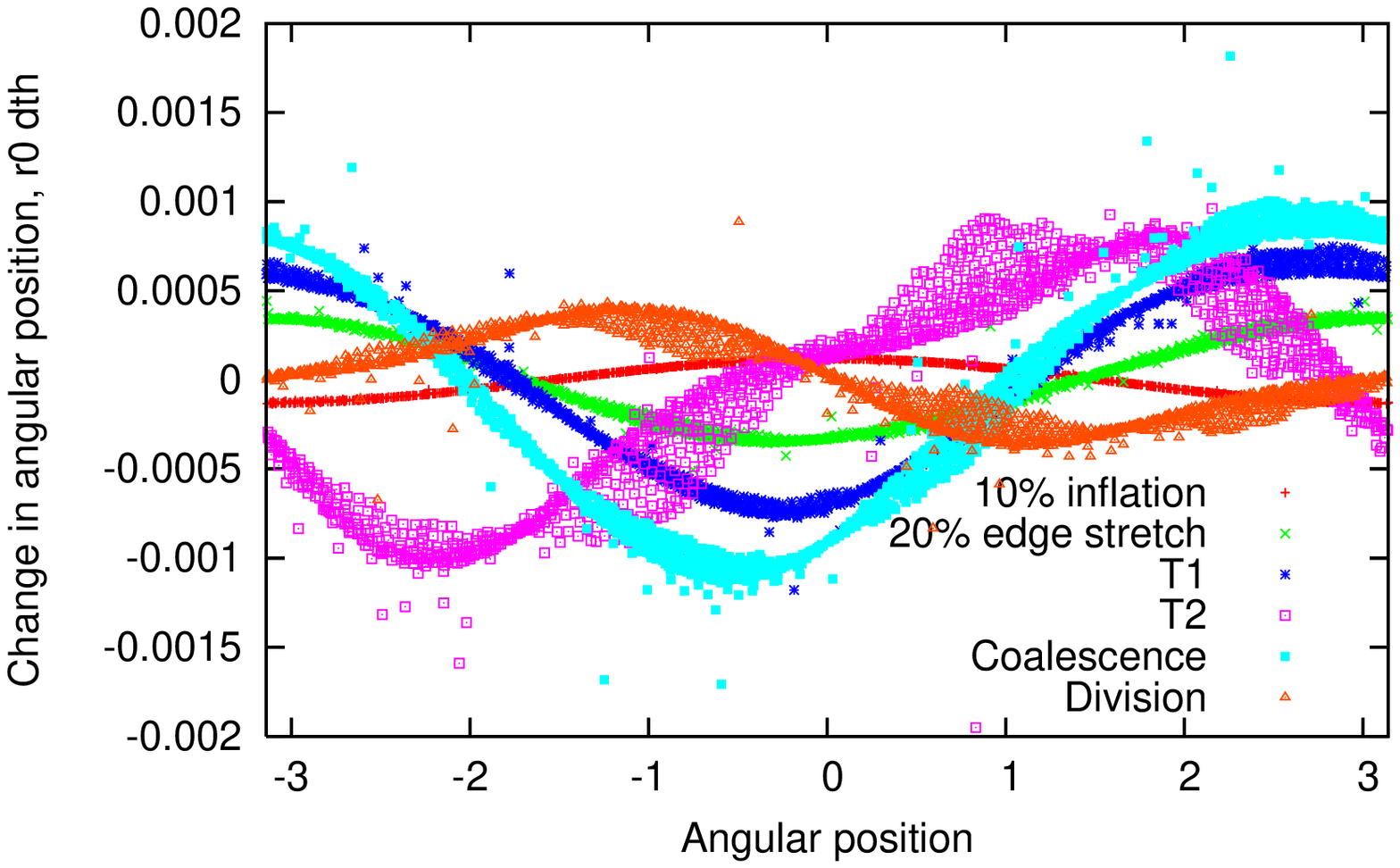}
}
\caption{Bubble displacement, for the same data shown in figure \ref{fig:displacement}.
(a) The magnitude of the radial displacement as a function of radial position.
(b) The radial displacement as a function of angular position.
 (c) The angular displacement as a function of angular position.}
\label{fig:deltar}
\end{center}
\end{figure}

\begin{figure}
\psfrag{Change in angular position, r0 dth}{Change in angular position, $r_0 \Delta \theta$}
\begin{center}
\mbox{
\subfigure[]{
\includegraphics[width=7cm,angle=270]{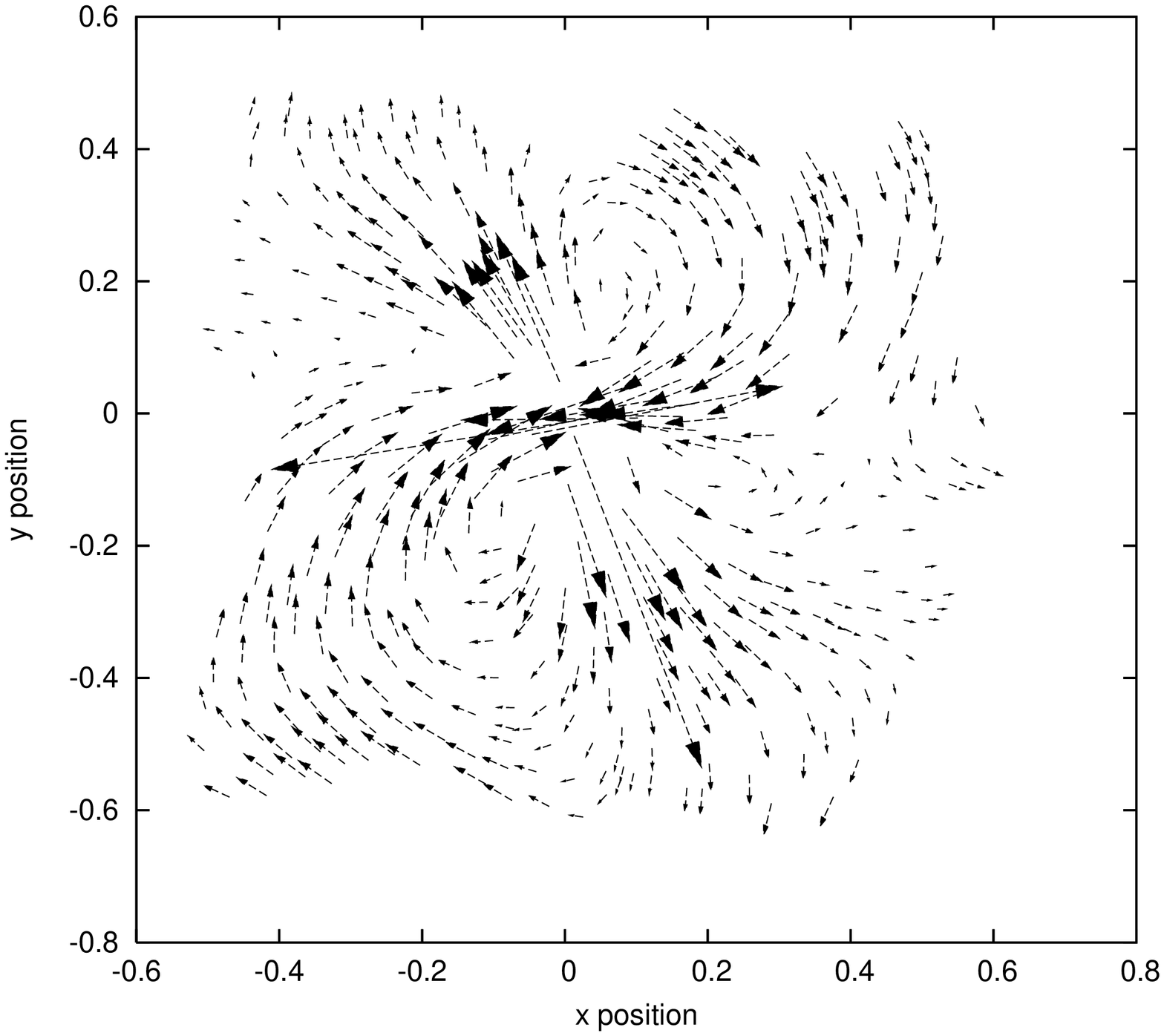}
}
\quad
\subfigure[]{
\includegraphics[width=7cm,angle=270]{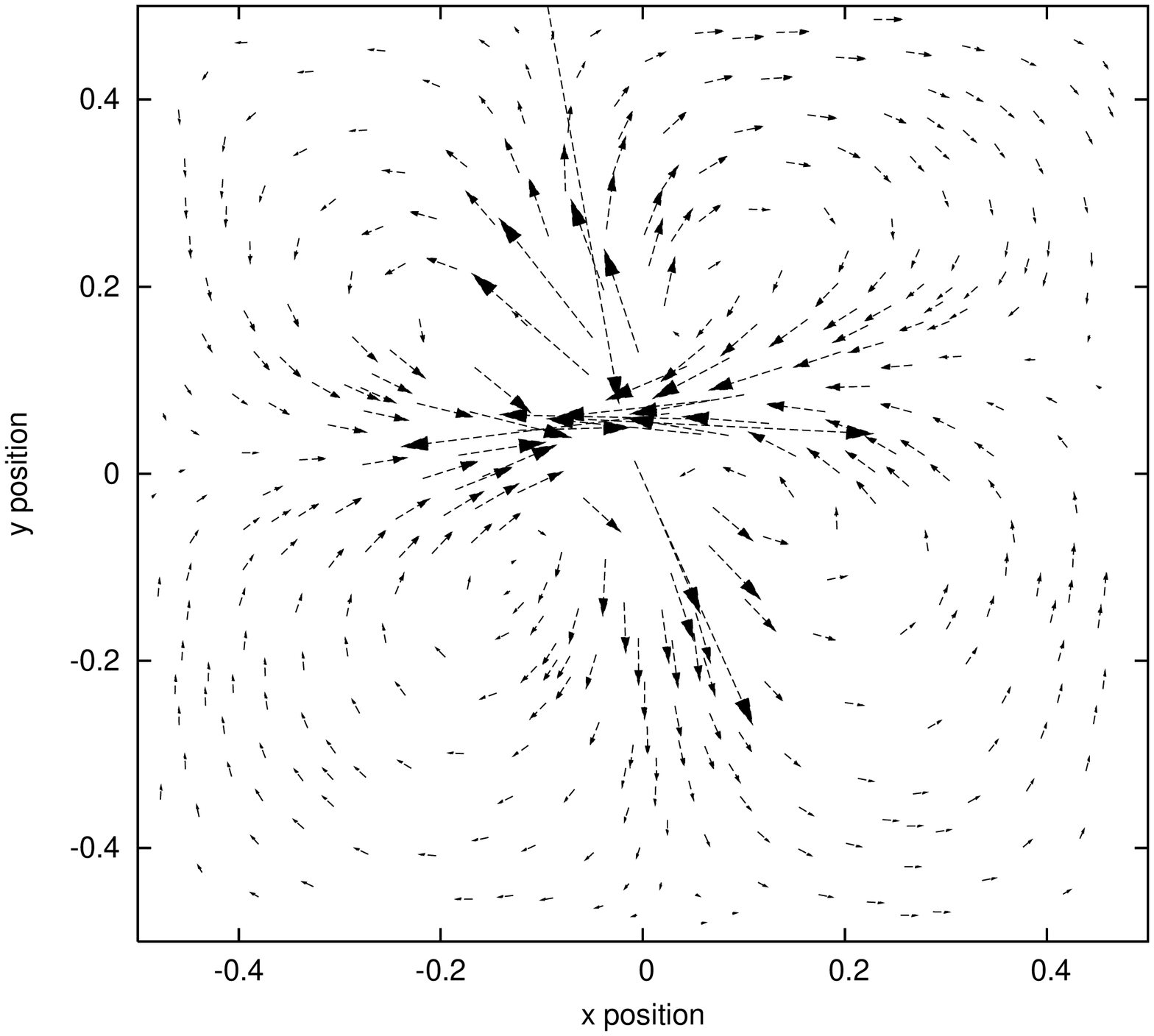}
}}
\mbox{
\subfigure[]{
\includegraphics[width=5cm,angle=270]{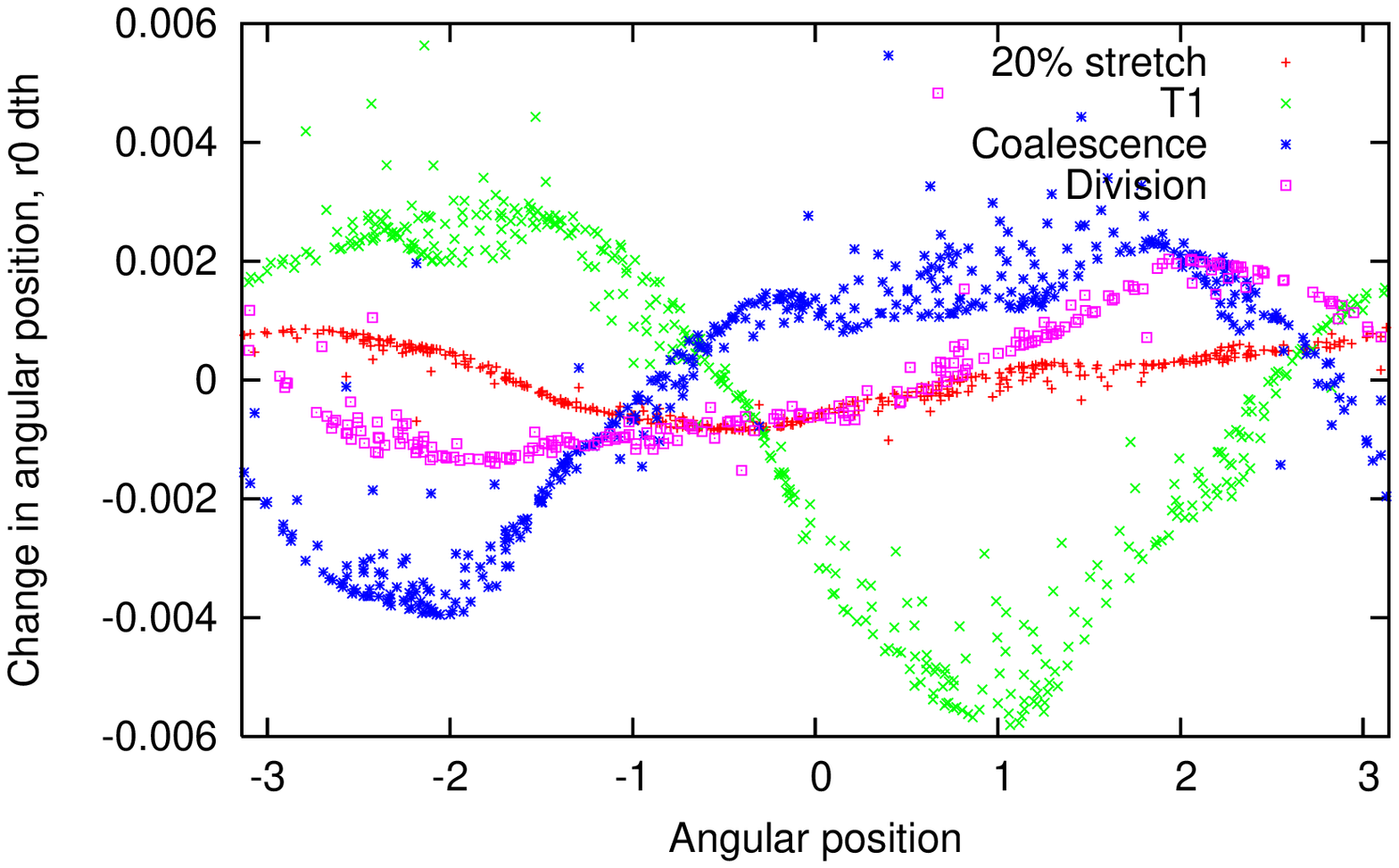}
}\quad
\subfigure[]{
\includegraphics[width=5cm,angle=270]{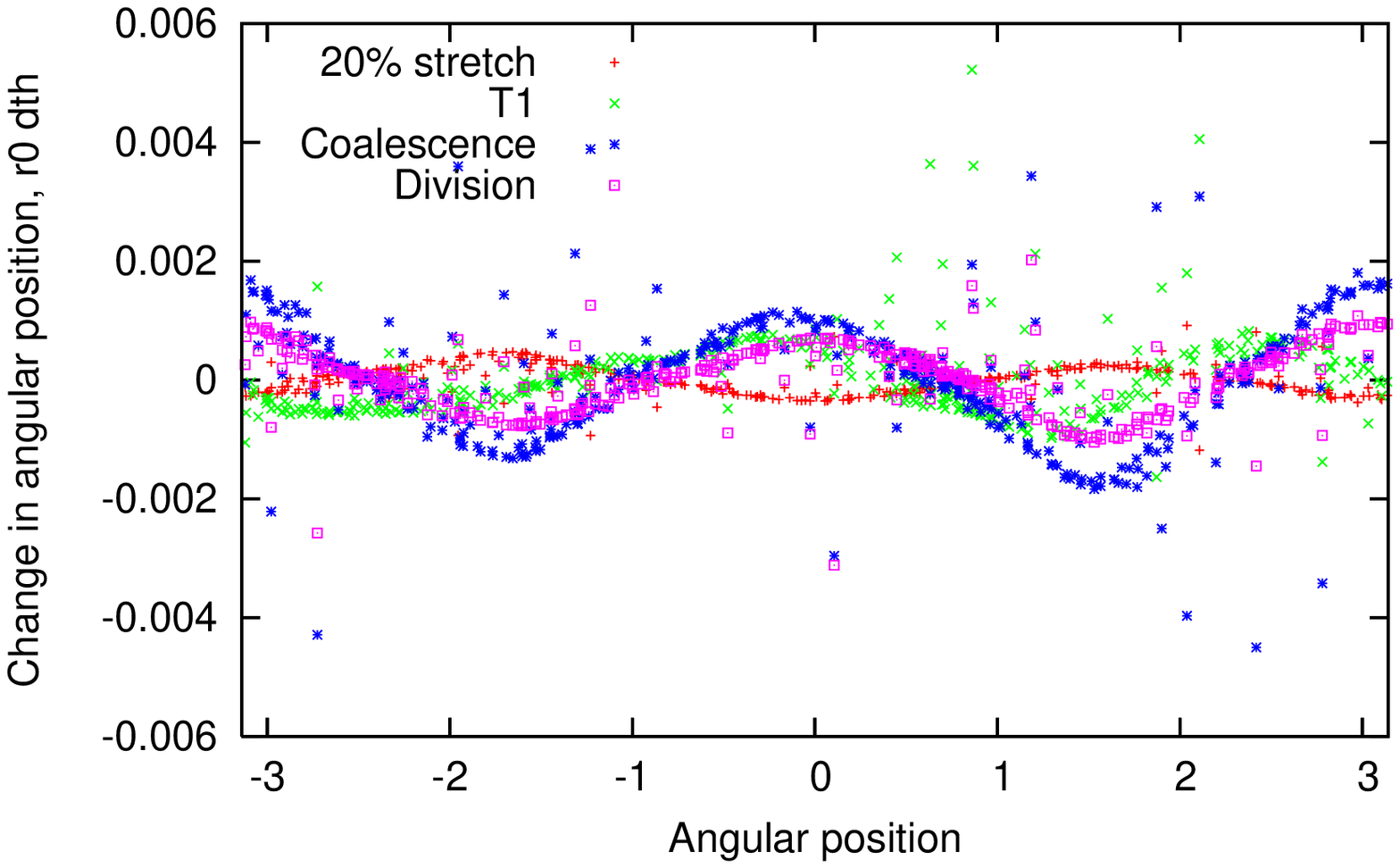}
}}
\caption{Effect of boundary conditions. Bubble displacement field  after a  T1 in a periodic foam of 400 bubbles
(a) and in a confined foam of 383 bubbles (b).
The average displacement and rotation have been subtracted, and the vector length multiplied by a factor of 30.
Angular bubble displacements in the same periodic (c) and confined (d) foams.
}
\label{fig:boundaries}
\end{center}
\end{figure}

\begin{figure}
\begin{center}
\includegraphics[width=12cm,angle=270]{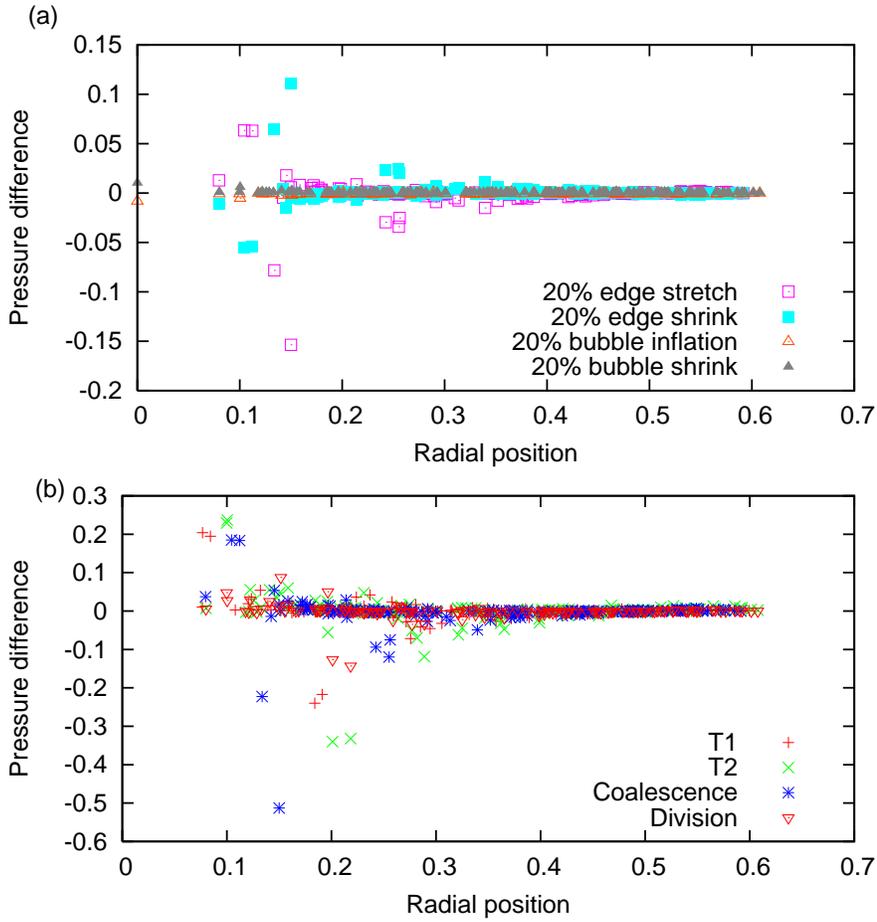}
\caption{The effect on the bubble pressures of each of the distortions in figure \ref{fig:subpics} applied to a disordered monodisperse cluster of 150 bubbles. On linear axes the effect of (a) an infinitesimal perturbation  is much less than after (b) the discontinuous processes (note the different scales). }
\label{fig:randmonopr}
\end{center}
\end{figure}

\begin{figure}
\psfrag{dp_sd}{$std(\Delta p)$}
\begin{center}
\subfigure[]{
\includegraphics[width=6cm,angle=270]{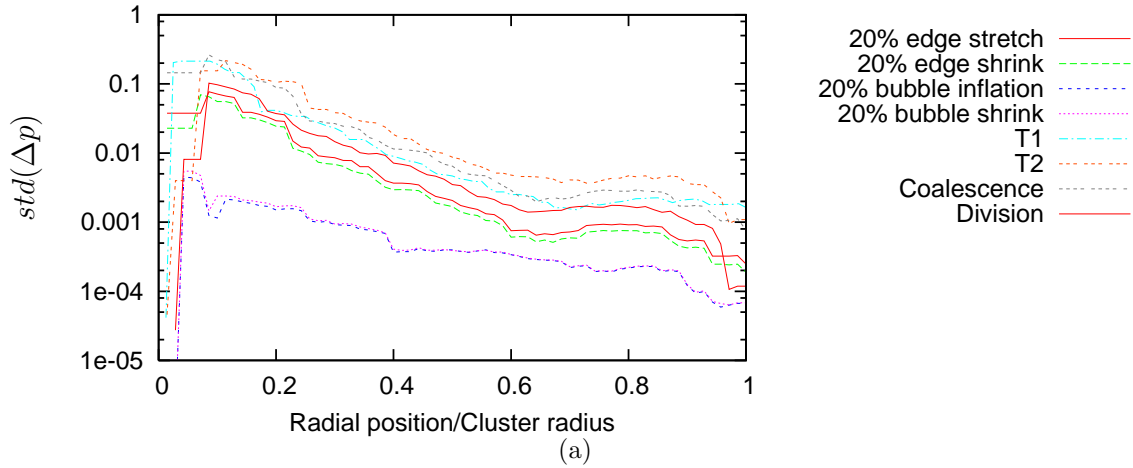}
}
\subfigure[]{
\includegraphics[width=11cm,angle=0]{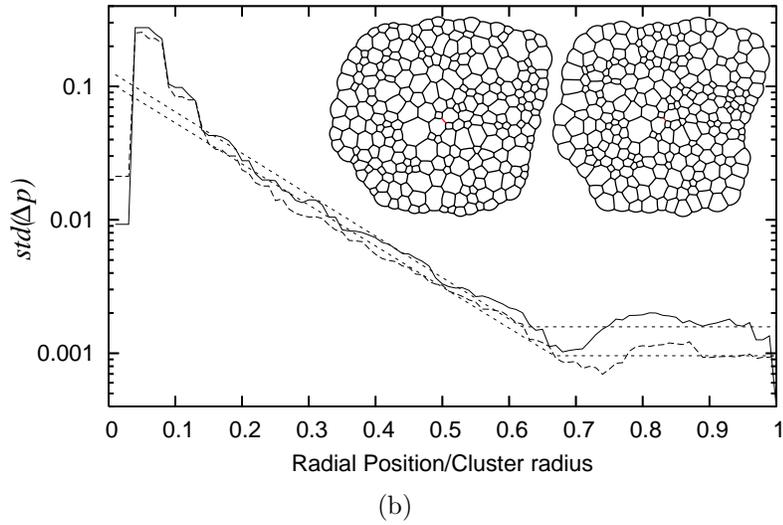}
}
\caption{
(a) Semi-log plot of the standard deviation of pressure change (see text)   against radial position measured in units of cluster radius. Same data as in figure \ref{fig:randmonopr}.
The drop-off in the data at large distance is caused by inaccuracy in defining the radius of the periphery of the cluster.
(b) Same plot for two highly polydisperse clusters of 250 bubbles. The dashed lines are fits to a piecewise linear function. The difference between the clusters is that the bubble volumes beyond eight bubble diameters from the centre have been randomly varied, as shown in the inset images of the clusters. This doesn't make a significant difference to the screening length.
}
\label{fig:outer}
\end{center}
\end{figure}

\begin{figure}
\psfrag{Disorder var(A)}{Disorder $var(A)$}
\psfrag{Inverse slope}{ $\ell_s/D$}
\centerline{
\includegraphics[width=11cm,angle=270]{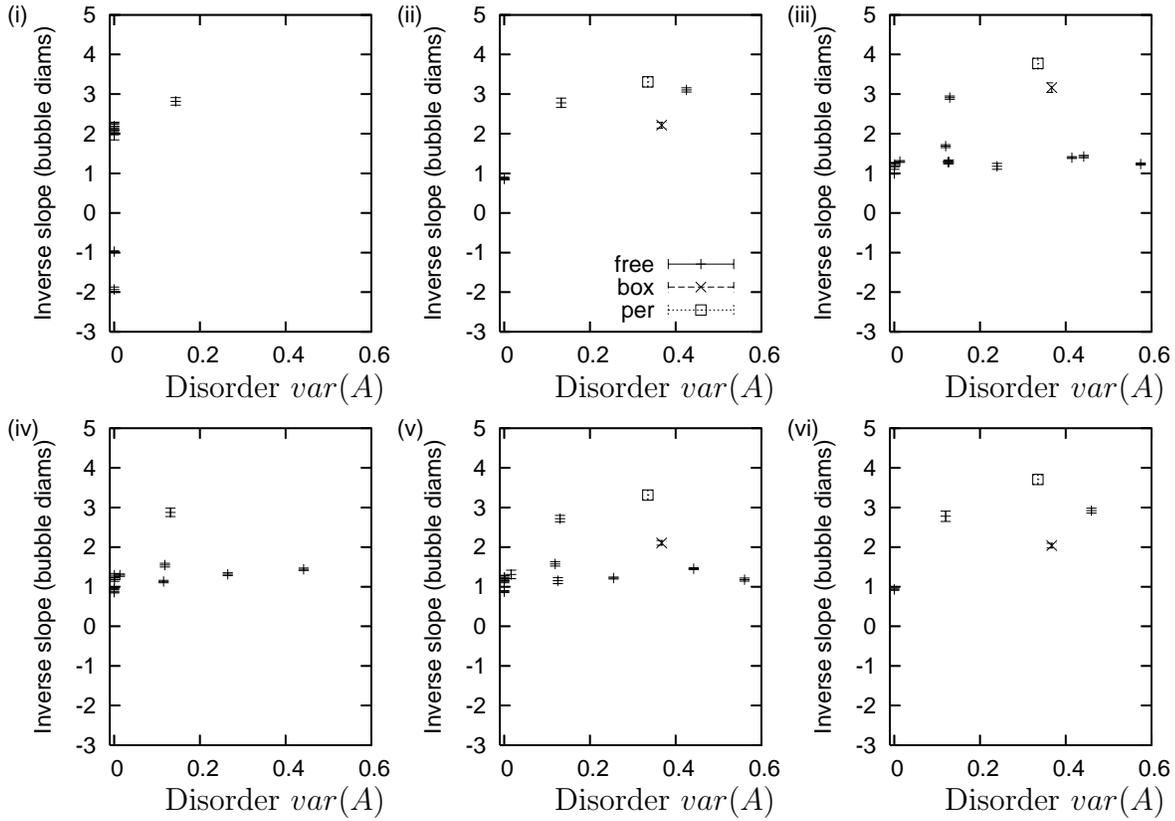}
}
\caption{Screening length $\ell_s$  of $std(\Delta p)$ (eq. \ref{eq:slope}), in units of bubble diameter  $D=2\sqrt{\langle A \rangle/\pi}$ , after (i) inflation (ii) an edge stretch, (iii) a T1, (iv) a T2, (v) coalescence, and (vi) division.}
\label{fig:alldata2}
\end{figure}

\begin{figure}
\psfrag{Disorder var(A)}{Disorder $var(A)$}
\psfrag{Critical radius}{ ${{r}}_c$ / foam radius}
\centerline{
\includegraphics[width=11cm,angle=270]{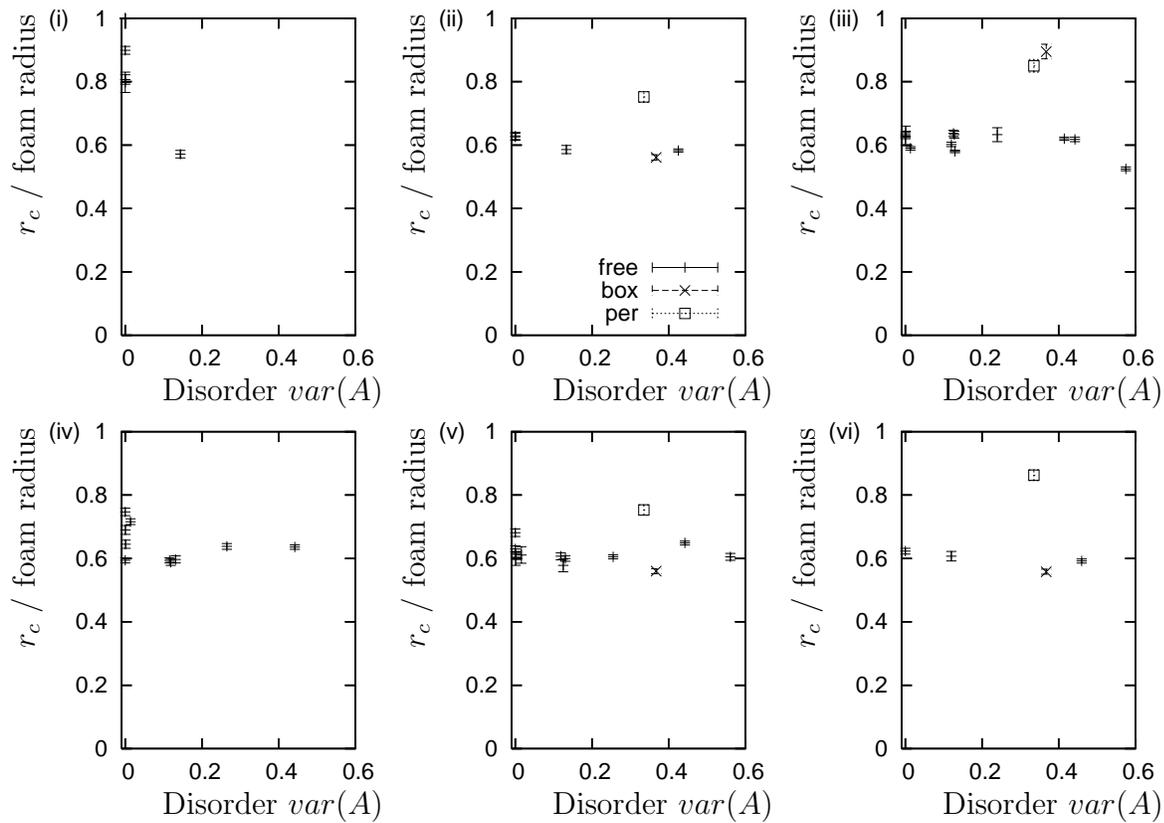}
}
\caption{Cross-over radius ${{r}}_c$, in units of foam radius, after (i) inflation, (ii) an edge stretch, (iii) a T1, (iv) a T2, (v) coalescence, and (vi) division. It is rarely possible to measure it for bubble inflation.}
\label{fig:alldata1}
\end{figure}

\begin{figure}
\psfrag{1400, dp}{1440, $\Delta p$}
\psfrag{1400, de}{1440, $\Delta e$}
\psfrag{1400, dl}{1440, $\Delta \ell$}
\psfrag{1261, dp}{1261, $\Delta p$}
\psfrag{1261, de}{1261, $\Delta e$}
\psfrag{1261, dl}{1261, $\Delta \ell$}
\centerline{
\includegraphics[width=8cm,angle=270]{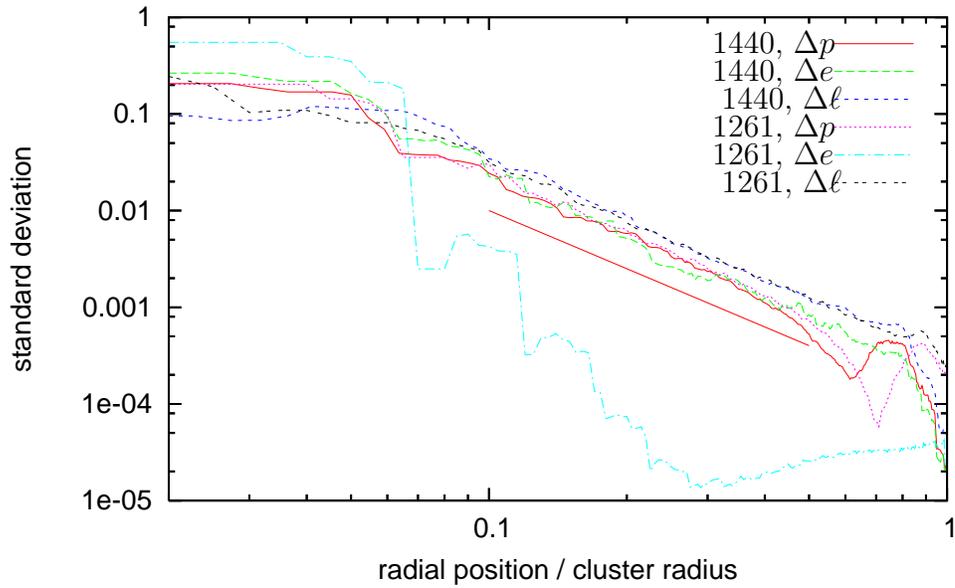}
}
\caption{
Log-log plot of the standard deviation of pressure difference $\Delta p$, perimeter change $\Delta e$ and edge-length differences $\Delta \ell$ against radial position measured in units of cluster radius, comparing data for a T1 in a disordered cluster of 1400 bubbles with a T1 in an ordered cluster of 1261 bubbles. In both cases, $\Delta \ell$ decreases quadratically (the solid line has slope -2) over the whole foam. $\Delta e$ does the same in the disordered cluster, but the ordered case shows a more rapid drop, then little change over the outer part of the cluster. The only scalar measure of disorder that shows screening in both ordered and disordered foams is the pressure change.
}
\label{fig:otherscalar}
\end{figure}

\begin{figure}
\psfrag{dp_sd}{$std(\Delta p)$}
\begin{center}
\subfigure[]{
\includegraphics[width=10cm,angle=270]{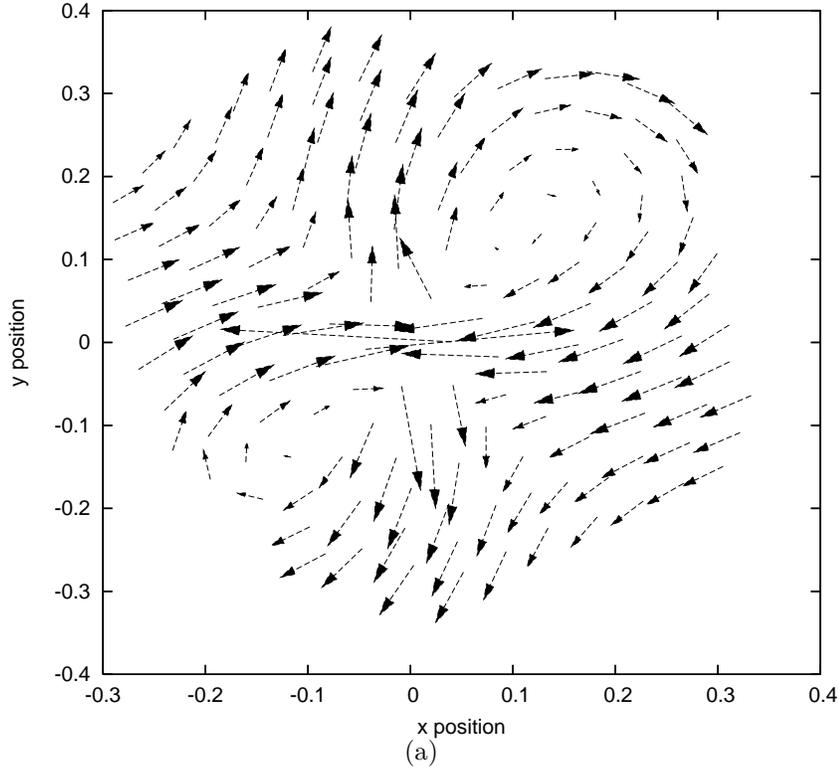}
}
\subfigure[]{
\includegraphics[width=10cm,angle=270]{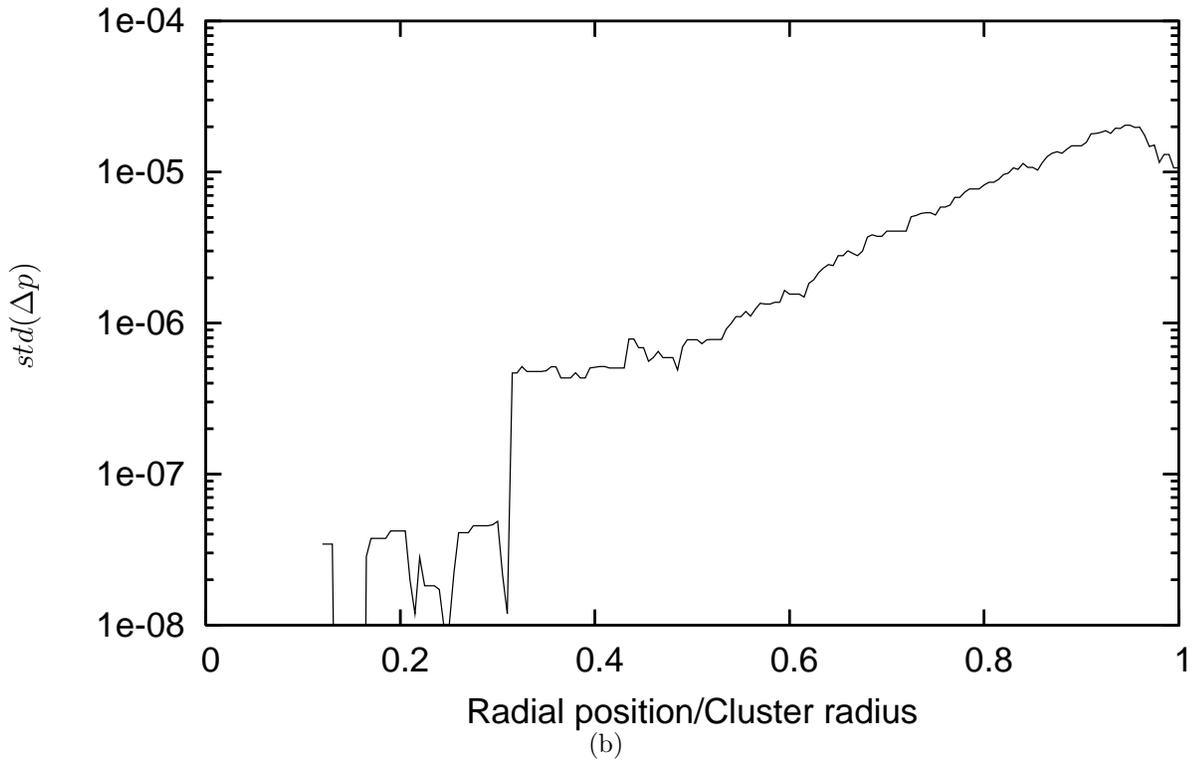}
}
\caption{The effect on the bubble pressures of a 20\% inflation event in a monodisperse ordered cluster of 1261 bubbles. In contrast to other perturbations to a monodisperse foam, and inflation events in disordered foams, there is no pressure screening: the pressure difference increases slightly towards the outside of the cluster. Note the scale, orders of magnitude smaller than figure \ref{fig:outer}. }
\label{fig:monoord1261pr}
\end{center}
\end{figure}

\begin{list}{}{\leftmargin 2cm \labelwidth 1.5cm \labelsep 0.5cm}

\item[\bf Fig. 1] Representative images of the foams used. Top row: Monodisperse foams, both ordered (217 and 1261 bubbles) and disordered (150 bubbles). Middle row: Polydisperse foams, in each case with almost the same value of polydispersity, $var(A) \approx 0.12$, and 250, 450 and 1400 bubbles respectively. Bottom row: Polydisperse foams without boundaries (periodic, 400 bubbles) and with boundaries (383 bubbles) respectively.

\item[\bf Fig. 2] The six types of distortion applied to the foams, illustrated here for the disordered monodisperse foam of 150 bubbles. Edge stretch (by 20\%), inflation (by 20\%), T1, T2, coalescence, division.

\item[\bf Fig. 3] The displacement field after (a) 10\% inflation, (b) 20\% edge stretch, (c) T1, (d) T2, (e) coalescence, and (f) division, in a cluster of 1400 bubbles with $var(A) \approx 0.12$. The average displacement and rotation has been subtracted in each case, and the vector length multiplied by a factor of 200 (in a), 100 (in b) or 30 (in c,d,e,f).

\item[\bf Fig. 4] Bubble displacement, for the same data shown in figure \ref{fig:displacement}.
(a) The magnitude of the radial displacement as a function of radial position.
(b) The radial displacement as a function of angular position.
 (c) The angular displacement as a function of angular position.

\item[\bf Fig. 5]  Effect of boundary conditions. Bubble displacement field  after a  T1 in a periodic foam of 400 bubbles
(a) and in a confined foam of 383 bubbles (b).
The average displacement and rotation have been subtracted, and the vector length multiplied by a factor of 30.
Angular bubble displacements in the same periodic (c) and confined (d) foams.

\item[\bf Fig. 6] The effect on the bubble pressures of each of the distortions in figure \ref{fig:subpics} applied to a disordered monodisperse cluster of 150 bubbles. On linear axes the effect of (a) an infinitesimal perturbation  is much less than after (b) the discontinuous processes (note the different scales). 

\item[\bf Fig. 7] (a) Semi-log plot of the standard deviation of pressure change (see text)   against radial position measured in units of cluster radius. Same data as in figure \ref{fig:randmonopr}.
The drop-off in the data at large distance is caused by inaccuracy in defining the radius of the periphery of the cluster.
(b) Same plot for two highly polydisperse clusters of 250 bubbles. The dashed lines are fits to a piecewise linear function. The difference between the clusters is that the bubble volumes beyond eight bubble diameters from the centre have been randomly varied, as shown in the inset images of the clusters. This doesn't make a significant difference to the screening length.

\item[\bf Fig. 8] Screening length $\ell_s$  of $std(\Delta p)$ (eq. \ref{eq:slope}), in units of bubble diameter  $D=2\sqrt{\langle A \rangle/\pi}$ , after (i) inflation (ii) an edge stretch, (iii) a T1, (iv) a T2, (v) coalescence, and (vi) division.

\item[\bf Fig. 9] Cross-over radius ${{r}}_c$, in units of foam radius, after (i) inflation, (ii) an edge stretch, (iii) a T1, (iv) a T2, (v) coalescence, and (vi) division. It is rarely possible to measure it for bubble inflation.

\item[\bf Fig. 10] Log-log plot of the standard deviation of pressure difference $\Delta p$, perimeter change $\Delta e$ and edge-length differences $\Delta \ell$ against radial position measured in units of cluster radius, comparing data for a T1 in a disordered cluster of 1400 bubbles with a T1 in an ordered cluster of 1261 bubbles. In both cases, $\Delta \ell$ decreases quadratically (the solid line has slope -2) over the whole foam. $\Delta e$ does the same in the disordered cluster, but the ordered case shows a more rapid drop, then little change over the outer part of the cluster. The only scalar measure of disorder that shows screening in both ordered and disordered foams is the pressure change.

\item[\bf Fig. 11] The effect on the bubble pressures of a 20\% inflation event in a monodisperse ordered cluster of 1261 bubbles. In contrast to other perturbations to a monodisperse foam, and inflation events in disordered foams, there is no pressure screening: the pressure difference increases slightly towards the outside of the cluster. Note the scale, orders of magnitude smaller than figure \ref{fig:outer}. 

\end{list}

\clearpage

\clearpage

\clearpage

\clearpage

\clearpage

 \clearpage

\clearpage

\clearpage

\clearpage

\clearpage

\end{document}